\documentclass[aps,pra,superscriptaddress,twocolumn]{revtex4-2}

\usepackage{graphicx}
\usepackage[dvipsnames]{xcolor}
\usepackage{bm}
\usepackage{physics}
\usepackage{amsthm}
\usepackage{mathtools}
\usepackage{hyperref}
\usepackage{bbm}
\usepackage{subcaption}
\usepackage{amssymb}
\usepackage[T1]{fontenc}
\newcommand{\valerio}[1]{{\color{red} #1}}

\newcommand{\ba}{\begin{eqnarray}}\newcommand{\ea}{\end{eqnarray}}\newcommand{\ban}{\begin{eqnarray*}}\newcommand{\ean}{\end{eqnarray*}}

\begin{document}

\title{Kochen-Specker nonlocal hidden variables must include time-ordering to allow for measurement independence of several agents}

\author{Valerio Scarani}\email{physv@nus.edu.sg}
\affiliation{Centre for Quantum Technologies, National University of Singapore, 3 Science Drive 2, Singapore 117543}
\affiliation{Department of Physics, National University of Singapore, 2 Science Drive 3, Singapore 117542}

\author{Antoine Suarez}\email{suarez.antoine@gmail.com}
\affiliation{\href{https://iis-edu.org/center-for-quantum-philosophy/}{Center for Quantum Philosophy}, Ackermanstrasse 25, CH-8044 Zürich   }

\date{\today}

\begin{abstract}
We consider an ontology, in which contextual nonlocal hidden variables are stored as pre-existing possibilities in a repository outside space-time; and in which the context can be chosen ``freely'' (measurement independence) by each agent, both in spacelike and timelike configurations. We show that, in Bell-type experiments involving several agents, for this ontology to be consistent, the context must include not only the measurements that can be performed, but also the time ordering of the choices of different agents.
\end{abstract}

\maketitle

\section{Introduction}

While the predictions of quantum theory are unchallenged, the description of ``how nature does it'' is still the object of speculations. In this note, we highlight a constraint on one such speculation that could be called \textit{ontological contextuality}. This interpretation is built on two key elements. First, as the name hopefully suggested, the hidden variables of Kochen and Specker (KS) \cite{KS} are given ontological status by assuming that all their possible values for every experiment are stored in a \textit{repository}. The second key element is to guarantee the \textit{``free choice''} of the context (measurement). When such a choice is made, the corresponding value is retrieved, the others turning into unused possibilities. By Bell's theorem, any hidden variable model should be nonlocal (for this and subsequent well known remarks on nonlocality, we refer the reader to \cite{RevModPhys.86.419,scarani}): so the repository should be thought as being outside spacetime.

This ontology can be detected in some of Specker's works, connected with his concerns as a Christian theologian. Already in 1960, seven years before KS, he brought the logic that would underpin quantum contextuality in connection with the theological question of the compatibility of divine omniscience and human free-will \cite{specker1960}. Citing from the English translation by M.P.~Seevinck: ``In a certain sense the scholastic speculations about the `Infuturabilien' also belong here, that is, the question whether the omniscience of God also extends to events that would have occurred in case something would have happened that did not happen.'' Here, `Infuturabilien' is a term invented by Specker, not translated by Seevinck and to be understood as something like `future contingencies'; the Scholastic debates to which reference is made are those sparked by Luis de Molina in the 16th century and hence known as Molinism.

One of us has recently argued that Specker's intuition can be turned into a theory of ``All-Possible-Experiments'' (APEX) \cite{suarez2019,suarez2025}, an interpretation that merges key intuitions of other interpretations. On the one hand, this interpretation has hidden variables like Bohm, but they are not dynamically generated by a quantum potential. Like the Many-Worlds interpretation, it gives some form of reality to all possible physical processes, but as pre-existing possibilities that the agents actualize in their single world, rather than as evolving splits that involve the agents as well.



Before continuing, let us pause to comment on related endeavors. Surely, ontological contextuality is not the only way to understand or use quantum contextuality. KS contextuality can be used as a tool in quantum foundations and quantum information without committing to an ontological reading \cite{RevModPhys.94.045007}. There are also notions of contextuality that differ from the KS narrative of "hidden" contextual variables, be they motivated operationally \cite{PhysRevA.71.052108} or ontologically \cite{grangier}. We leave open whether the observation we make here is relevant to some of those other views as well. Meanwhile, in the interpretations of statistical physics, Albert has proposed to consider probability distributions over all possible worlds, an interpretation later called the ``Mentaculus'' by reference to a cult movie by the Coen brothers \cite{Loewer2020-LOETMV-2}. To the best of our knowledge, the Mentaculus has not been explicitly formulated for quantum mechanics and is not concerned with free choice of agents.

Back to our main track: central to what we are discussing is the \textit{desideratum} of ``free choice'' of the context. A libertarian view of human free will was certainly at the back of Specker's mind, as is of ours. But all we are going to say requires only a much weaker \textit{measurement independence}: the processes that select the context should be uncorrelated from (``random for'') the box that prepares the state. This independence may be argued for in various ways and does not need to any creative, ``absolutely random'' (whatever that may mean) process. For instance, one may believe that our free is illusory, and still maintain that our mental processes and the processes that prepare the state are independent. Or, like most experiments do, some devices in the lab can be trusted to be random number generators. Even less may be acceptable, down to choosing the measurements of a Bell experiment using the text of some social media posts \cite{pironio}: while these may have been generated long ago and are public, assuming that the source of entangled particles is correlated with those posts leads to a worldview that is very close (if not identical) to superdeterminism. For our purposes, it does not matter which of these options is chosen to affirm measurement independence. It is also known that measurement independence can be relaxed to some extent when the goal is to certify Bell nonlocality; but our goal here is to describe a universe in which complete measurement independence \textit{can be the case} and must therefore be accounted for.

\section{The observation}

Within this framework, our observation is the following: \textit{for experiments where the measurement is selected by two or more independent processes, if measurement independence has to hold, the time ordering of these processes should be part of the context as well}.

\subsection{A simple example}

This is better introduced with the simplest example: the violation of the CHSH inequality $\sum_{x,y\in\{0,1\}}(-1)^{a+b-xy}\leq 2$. Since every unrestricted probability distribution can always be seen as a mixture of deterministic assignement, there is no loss of generality in assuming that the repository contains one such assignment for each run of the test. The nonlocal deterministic assignments that violate such an inequality are such that $(a+b)\textrm{mod}2=xy$, for instance
\begin{align}
    \begin{array}{cc|cc}
        x & y & a & b  \\\hline
        0 & 0 & 0 & 0  \\
        0 & 1 & 0 & 0  \\
        1 & 0 & 0 & 0  \\
        1 & 1 & 0 & 1  
    \end{array}\label{1sign}
\end{align} where on the left of the vertical line we have the context, and on the right the value of the hidden variable. This example is signaling from Alice to Bob: by choosing input $y=1$, Bob gets the output $b=x$ and thus learns Alice's input. Now, suppose that in a given round this is the contextual assignment stored in the repository; and that Bob measures with input $y=1$ and sees output $b$ well before Alice performs her measurement: then Alice would be obliged to input $x=b$, and is no longer free. This argument is general because all the deterministic assignments that are not local are signaling. We note that Colbeck and Renner made the same observation with a more operational narrative: if a shared resource describes a fixed signaling distribution, then free choice of inputs cannot be enforced \cite{CR11}. For our narrative, we are going to read the implication in reverse: we want to ensure measurement independence, but the repository must contain signaling distributions; so the repository must contain more than just one signaling assignment.

As announced, the most straightforward way to fix the problem is to \textit{add time ordering to the context}. For the example above, the upgraded context on the left of the line would be:
\begin{align}
    \begin{array}{ccc|cc}
        x & y & \textrm{time order}& a & b  \\\hline
        0 & 0 & \textrm{any}& 0 & 0  \\
        0 & 1 & \textrm{any}& 0 & 0  \\
        1 & 0 & \textrm{any}& 0 & 0  \\
        1 & 1 & t_A\leq t_B& 0 & 1  \\
        1 & 1 & t_A> t_B& 1 & 0  
    \end{array}\,,
\end{align}
This fix applies both in the case of an absolute time and in the relativistic case; in the latter, the condition $t_A> t_B$ should be understood as ``the event of Alice is in the future light cone of that of of Bob''.

\subsection{Comments on time ordering and measurement independence}

Now that the gist of our argument has been spelled out, before making it general, we must address a few clarifications.

The time ordering we are referring to here is the ordering between the choices of measurements by \textit{several independent agents}. We stress it because the literature on contextuality often focuses on single systems, and a contextuality test may feature sequential measurements by a single agent. In that case, the agent should be free to choose the order of the measurement, but the second choice does not need to be independent from the first (fortunately, as it would be quite hard to justify).

Of course, in a timelike separated experiment, the information about Bob's earlier choice and outcome \textit{could} have reached Alice: no overarching law of physics forbids her loss of measurement independence. The mathematics of such a view have been presented by Horodecki and Ramanathan \cite{HR19}. Our reply to such an objection is that the collapse of independence for timelike-separated agents is neither what we experience for human free will, nor is what is believed to be the case for quantum random number generators, and is even less the case if we accept independence between the quantum source and a list of social media posts. In this sense, we are using a stronger form of measurement independence than the one needed only for spacelike separation.

Finally, underlying this whole discussion is the question: what events are we looking at? When discussing the locality loophole, the relevant times are the choice of the input of one player and the production of the output of the other (as well known, both are hard to pinpoint, and experiments should resort to operational definitions). However, in an ontology where free choices are possible, ``It ain't over till it's over'' \footnote{Yogi Berra, interview, July 1973.}: the systems under study do not know what experiment they are undergoing till they meet the measurement device or procedure. It is at that point in time that they have to consult the repository for the outcome to give to that particular experiment. Particularly relevant for our discussion is the possibility that one player simply delays the measurement. So, in the table above and in all our considerations, the time $t_P$ should be understood as the time when the system in the hands of Player produces its outcome. It remains one of the events that are hard to pinpoint for us humans; but in discussing this single-world ontology, all that matters is that such a time exists and the context in the repository should depend on it. Interestingly, this is decoupled from the debate on Wigner's friends and the absoluteness of outcomes. If the context is defined without time ordering as in Eq.~\eqref{1sign}, it is a fact that Alice loses her freedom if $t_{A}> t_{B}$, even if at a time $t_W>t_{A}$ Wigner ``hadamards Bob's brain'' and Bob's outcome is forever forgotten.

\subsection{General construction}

Next, we need to show that such a fix is always possible; in particular, for multipartite quantum correlations, where the possible time-orderings are increasingly complex. It may be obvious to the reader, especially thinking of Bohmian mechanics; and eventually it will prove to be; but it is better to spell out explicitly what could go wrong.

Among the nonlocal deterministic assignments, some have circular signaling. In the bipartite case, an assignment like
    \begin{align}
    \begin{array}{cc|cc}
        x & y & a & b  \\\hline
        0 & 0 & 0 & 0  \\
        0 & 1 & 1 & 0  \\
        1 & 0 & 0 & 1  \\
        1 & 1 & 1 & 1  
    \end{array}\label{2sign}
\end{align} is sending Bob's input to Alice ($a=y$) and Alice's input to Bob ($b=x$). If the repository contains one such assignment, the fix described previously could not work: in a timelike experiment, at least one of the parties would lose their independence. For the CHSH example we have been using so far, it is rather immediate that assignments of the type \eqref{2sign} are not needed. But we need to prove that \textit{every quantum correlation $P(a,b,c,...|x,y,z,...)$ can be decomposed only on assignments with given time ordering,} i.e.~that have no signaling cycle. Further, such a decomposition should exist \textit{for every time ordering.} 

Fortunately, any quantum correlation can indeed be decomposed on assignments with given time ordering, for every ordering (in fact, the whole no-signaling polytope is contained in the so-called arrow-of-time polytope \cite{Hoffmann_2018}). The argument is rather simple: we write it for three parties, the generalisation being straightforward. Since quantum correlations are no-signaling, it holds
\begin{align}
P(a,b,c|x,y,z)=P(a|x)P(b|y,x,a)P(c|z,y,b,x,a) \label{pcond}
\end{align}
because $P(a|x,y,z)=P(a|x)$ and $P(a,b|x,y,z)=P(a,b|x,y)$. Each factor of \eqref{pcond} can then be decomposed into deterministic assignments: local assignments for the first; conditionals of assignments that signal from $A$ to $B$ for the second; conditionals of assignments that signal from $(A,B)$ to $C$ for the third. So \eqref{pcond} is the expansion in conditional probabilities that is suitable for time ordering $t_A\leq t_B\leq t_C$. For other time orderings, one just looks at the corresponding expansion. Notice that the decompositions of the form \eqref{pcond} needed here are different from those of interest in the study of multipartite nonlocality \cite{PhysRevA.88.014102}: there, the question is whether a given $P$ can be formally decomposed on assignments that involve only fewer-party signaling, but possibly in any direction, without reference to the time of the choice of the inputs. 



\section{Conclusion}

In conclusion, we have considered an ontology where contextual nonlocal hidden variables are given the status of pre-existing possibilities, one of the possibilities being later actualized by the choice of the context. We have shown that, in Bell-type experiments involving several independent agents, time-ordering must be part of the context if ``free choice'' of the context (in the sense of measurement independence discussed in the paper) has to hold for all the agents. 

This ontology may be considered one of the minimal explanations for how outcomes are produced in a quantum experiment, insofar as it requires a single universe and saves measurement independence (it is well known that more radical departures from these common beliefs have been proposed: it is not our purpose here to compare their merits). The main role in this ontology is played by the repository in which the possibilities are stored. Since hidden variable must be nonlocal to account for the violation of Bell inequalities, this repository should be outside space-time. This notion resonates with ontologies developed for other purposes than quantum physics, like God’s omniscient mind in the theological schools who inspired Ernst Specker, or the Mentaculus view of statistical physics. In the light of our analysis, it would be interesting to explore if the latter can be seen as a subset of our larger “quantum” repository consisting of experiments with free choices and nonlocal assignments. 


\section*{Acknowledgments}

Our output, though entirely free, was influenced by extensive inputs by Clive Aw and Costantino Budroni, as well as other inputs from Nicolas Brunner, Ad\'an Cabello, Nicolas Gisin, Philippe Grangier and Jef Pauwels.

V.S. is supported by the National Research Foundation, Singapore through the National Quantum Office, hosted in A*STAR, under its Centre for Quantum Technologies Funding Initiative (S24Q2d0009); and by the Ministry of Education, Singapore, under the Tier 2 grant ``Bayesian approach to irreversibility'' (Grant No.~MOE-T2EP50123-0002).

\bibliography{references}

@article{PhysRevA.88.014102,
  title = {Definitions of multipartite nonlocality},
  author = {Bancal, Jean-Daniel and Barrett, Jonathan and Gisin, Nicolas and Pironio, Stefano},
  journal = {Phys. Rev. A},
  volume = {88},
  issue = {1},
  pages = {014102},
  numpages = {5},
  year = {2013},
  month = {Jul},
  publisher = {American Physical Society},
  doi = {10.1103/PhysRevA.88.014102},
  url = {https://link.aps.org/doi/10.1103/PhysRevA.88.014102}
}

@article{RevModPhys.94.045007,
  title = {Kochen-Specker contextuality},
  author = {Budroni, Costantino and Cabello, Ad\'an and G\"uhne, Otfried and Kleinmann, Matthias and Larsson, Jan-\AA{}ke},
  journal = {Rev. Mod. Phys.},
  volume = {94},
  issue = {4},
  pages = {045007},
  numpages = {62},
  year = {2022},
  month = {Dec},
  publisher = {American Physical Society},
  doi = {10.1103/RevModPhys.94.045007},
  url = {https://link.aps.org/doi/10.1103/RevModPhys.94.045007}
}

@article{RevModPhys.86.419,
  title = {Bell nonlocality},
  author = {Brunner, Nicolas and Cavalcanti, Daniel and Pironio, Stefano and Scarani, Valerio and Wehner, Stephanie},
  journal = {Rev. Mod. Phys.},
  volume = {86},
  issue = {2},
  pages = {419--478},
  numpages = {60},
  year = {2014},
  month = {Apr},
  publisher = {American Physical Society},
  doi = {10.1103/RevModPhys.86.419},
  url = {https://link.aps.org/doi/10.1103/RevModPhys.86.419}
}

@article{CR11,
	author = {Colbeck, Roger and Renner, Renato},
	date = {2011/08/02},
	doi = {10.1038/ncomms1416},
	id = {Colbeck2011},
	isbn = {2041-1723},
	journal = {Nature Communications},
	number = {1},
	pages = {411},
	title = {No extension of quantum theory can have improved predictive power},
	url = {https://doi.org/10.1038/ncomms1416},
	volume = {2},
	year = {2011}}

@article{grangier,
  title={Revisiting the Interpretations of Quantum Mechanics: From FAPP Solutions to Contextual Ontologies},
  author={Grangier, Philippe},
  journal={arXiv preprint arXiv:2601.20488},
  year={2026}
}

@article{KS,
  title = {The Problem of Hidden Variables in Quantum Mechanics},
  author = {Kochen, Simon and Specker, Ernst P.},
  journal = {Journal of Mathematics and Mechanics},
  volume = {17},
  issue = {1},
  pages = {59-87},
  numpages = {29},
  year = {1967},
  month = {},
  publisher = {ndiana University Mathematics Journal}
}

@article{pironio,
  title={Random 'choices' and the locality loophole},
  author={Pironio, Stefano},
  journal={arXiv preprint arXiv:1510.00248},
  year={2015}
}

@book{scarani,
    author = {Scarani, Valerio},
    title = {Bell nonlocality},
    publisher = {Oxford University Press},
    year = {2019}
}

@article{PhysRevA.71.052108,
  title = {Contextuality for preparations, transformations, and unsharp measurements},
  author = {Spekkens, R. W.},
  journal = {Phys. Rev. A},
  volume = {71},
  issue = {5},
  pages = {052108},
  numpages = {17},
  year = {2005},
  month = {May},
  publisher = {American Physical Society},
  doi = {10.1103/PhysRevA.71.052108},
  url = {https://link.aps.org/doi/10.1103/PhysRevA.71.052108}
}

@article{suarez2025,
  title={Beyond Space and Time: Quantum Superposition as a Real-Mental State About Choices},
  author={Suarez, Antoine},
  journal={Condensed Matter},
  volume = {10},
  issue = {3},
  pages = {43},
  year = {2025},
  month = {August},
  publisher = {MDPI Open Access Journals},
  doi = {10.3390/condmat10030043},
  url = {https://doi.org/10.3390/condmat10030043}
}

@article{suarez2019,
  title={All-Possible-Worlds: Unifying Many-Worlds and Copenhagen, in the Light of Quantum Contextuality},
  author={Suarez, Antoine},
  journal={arXiv preprint arXiv:1712.06448v2},
  year={2019},
  url={https://doi.org/10.48550/arXiv.1712.06448}
}

@article{specker1960,
  title={Die Logik nicht gleichzeitig entscheidbarer Aussagen},
  author={Specker, Ernst},
  journal={Dialectica},
  volume = {14},
  pages = {239},
  year = {1960},
  url = {https://doi.org/10.48550/arXiv.1103.4537}
}

@incollection{Loewer2020-LOETMV-2,
	author = {Barry Loewer},
	booktitle = {Statistical Mechanics and Scientific Explanation: Determinism, Indeterminism and Laws of Nature},
	editor = {Valia Allori},
	publisher = {World Scientific},
	title = {The Mentaculus Vision},
	year = {2020}
}

@article{HR19,
	author = {Horodecki, Pawe{\l} and Ramanathan, Ravishankar},
	date = {2019/04/12},
	doi = {10.1038/s41467-019-09505-2},
	id = {Horodecki2019},
	isbn = {2041-1723},
	journal = {Nature Communications},
	number = {1},
	pages = {1701},
	title = {The relativistic causality versus no-signaling paradigm for multi-party correlations},
	url = {https://doi.org/10.1038/s41467-019-09505-2},
	volume = {10},
	year = {2019}}

@article{Hoffmann_2018,
	author = {Hoffmann, Jannik and Spee, Cornelia and G{\"u}hne, Otfried and Budroni, Costantino},
	doi = {10.1088/1367-2630/aae87f},
	journal = {New Journal of Physics},
	month = {oct},
	number = {10},
	pages = {102001},
	publisher = {IOP Publishing},
	title = {Structure of temporal correlations of a qubit},
	url = {https://doi.org/10.1088/1367-2630/aae87f},
	volume = {20},
	year = {2018}}

\end{document}